\documentclass[a4paper,xcolor=dvipsnames]{jpconf}
\usepackage{graphicx}
\usepackage{amsmath,amssymb,bbm}
\usepackage{amsfonts}
\usepackage{pgf}

\newcommand{\ro}{\hat{\rho}}
\newcommand{\Ho}{\hat{H}}
\newcommand{\qo}{\hat{q}}
\newcommand{\Po}{\hat{P}}
\newcommand{\Uo}{\hat{U}}
\newcommand{\Co}{\hat{C}}

\newcommand{\bra}[1]{\langle#1\vert}
\newcommand{\ket}[1]{\vert#1\rangle}
\newcommand{\I}{\mathrm{i}}

\begin{document}
\title{Classical-Quantum Coexistence: a `Free Will' Test}

\author{Lajos Di\'osi}

\address{Wigner Research Center for Physics, H-1525 Budapest 114, P.O.Box 49, Hungary}

\ead{diosi@wigner.mta.hu}

\begin{abstract}
Von Neumann's statistical theory of quantum measurement interprets 
the instantaneous quantum state and derives instantaneous classical variables. 
In realty, quantum states and classical variables coexist and can influence each 
other in a time-continuous way. This has been motivating investigations 
since longtime in quite different fields from quantum cosmology to optics as well 
as in foundations. Different theories (mean-field, Bohm, 
decoherence, dynamical collapse, continuous measurement, hybrid dynamics, e.t.c.) 
emerged for what I call `coexistence of classical continuum with quantum'. 
I apply to these theories a sort of `free will' test to distinguish
`tangible' classical variables useful for causal control from useless ones. 
\end{abstract}

\section{Introduction}\label{Introduction}
It is now widely agreed that the fundamental description of physical systems is 
the quantum theory. The state of the given closed system is described by the density matrix
$\ro$, the Hamiltonian $\Ho$ governs its dynamics by the von Neumann equation of motion:
\begin{equation}\label{vN}
\frac{d\ro}{dt}=\frac{-\I}{\hbar}[\Ho,\ro]\;.
\end{equation}
This yields the unitary evolution 
\begin{equation}\label{UroU}
\ro\longrightarrow\Uo(t)\ro\Uo^\dagger(t)
\end{equation}
where $\Uo(t)$ solves $i\hbar d\Uo/dt=\Ho\Uo$ with the initial condition $\Uo(0)=\hat 1$.
The density matrix $\ro$ is interpreted statistically by the concept of measurement \cite{Neu32}. 
The role of the latter is that it assigns classical variables $z$ to the quantum 
density matrix $\ro$:
\begin{equation}\label{ro-z}
\ro\Longrightarrow z\;.
\end{equation}
Without such assignment, the quantum theory would be incomplete because the matrix $\ro$ 
itself cannot be identified as classical variables.
A matrix element, say $\rho_{11}$, of $\ro$ does not behave like true
classical variables should. How should true classical variables behave?
The answer is a new `Free Will Test' to distinguish the formal classical variables 
from the true ones which we call \emph{tangible}.

\section{Free will}\label{Free}
Suppose we have assigned a classical variable $z$ to the quantum state $\ro$ 
at a given time. We expect of a tangible classical $z$ that we can use it at our free will 
to control the future dynamics of our system. We expect in particular that
we can make the future unitary evolution depend on $z$.   
If the resulting theory remains consistent then $z$ is tangible classical variable and
the given assignment $\ro\Rightarrow z$ passes the Free Will Test. 

FWT is satisfied within standard quantum theory. As we said, 
the assignment $\ro\Rightarrow z$ is given by a von Neumann measurement. The  
mathematical ingredient is a complete orthogonal set of projectors $\Po_z$ where the
labels $z$, also called measurement outcomes, are the possible classical values 
that will be assigned to $\ro$ at random with probability $p_z=\tr(\Po_z\ro)$, respectively. 
The emergence of the classical variable $z$ is accompanied by the instantaneous
`collapse' of the state:
\begin{equation}\label{coll}
\ro\longrightarrow \frac{1}{p_z}\Po_z\ro\Po_z\equiv\ro_z\;.
\end{equation}
Now, according to FWT, we assume a $z$-dependent unitary dynamics $\Uo_z$, evolve each 
$\ro_z$ and average them over the statistics $p_z$ of the classical variable 
$z$, yielding:
\begin{equation}\label{vNz_meas}
\ro\longrightarrow \ro^\prime =\sum_z p_z\Uo_z\ro_z\Uo_z^\dagger =\sum_z\Uo_z\Po_z\ro\Po_z\Uo_z^\dagger\;.
\end{equation}
We find that the map $\ro\rightarrow\ro^\prime$ is linear. This linearity is necessary \cite{Dio11}
for the statistical interpretation \cite{Neu32} of $\ro^\prime$, it maintains consistency of the
theory. Hence we see the measurement outcomes $z$ are tangible classical variables, FWT has been
passed by standard quantum mechanics.

FWT is powerful. In standard quantum mechanics it singles out quantum measurement theory
as the only way to assign tangible classical variables to the quantum state.
Measurement outcomes are the only tangible classical variables. This also implies
that a deterministic assignment $\ro\Rightarrow z$ will never pass FWT, will never yield a tangible $z$.
Had we assigned $z$ to the upper-left matrix element of
$\ro$ the resulting dynamics $\ro\rightarrow\ro^\prime$  would be non-linear which is a typical inconsistency
indicating that $z$ is not tangible . A frequently used non-tangible classical variable is the quantum 
mean, e.g., of the coordinate $\qo$:
\begin{equation}\label{z_mean}
z=\tr(\qo\ro)\;.
\end{equation}
For this deterministic way of assignment the $z$-dependent dynamics
\begin{equation}\label{UzroUz}
\ro\longrightarrow\ro^\prime =\Uo_z\ro\Uo_z^\dagger
\end{equation}
is non-linear and invalidates the statistical interpretation of $\ro^\prime$. 
Tangible classical variable is the Husimi assignment of phase-space coordinates $z\equiv(q,p)$ of a 
quantum harmonic oscillator \cite{Hus40}. It can be underlied by the measurement of the overcomplete 
set of coherent state projectors $\Po_z=\ket{z}\bra{z}$ . The probability of the assignment is $p_z=\tr(\Po_z\ro)$. 
The consistency (linearity) of the $z$-dependent dynamics follows from the same structure 
(\ref{coll}-\ref{vNz_meas}) as in case of standard projective measurements.

\section{Classical-quantum coexistence}\label{C-Q}
We experience quantum and classical phenomena in our surrounding. Classical ones 
are apparent, they emerge from the quantum ones according to quantum theory. 
This explanation can be extended for all phenomena
except for quantum gravity where the emergence of classical continuum is still
waiting for a consistent theory. For this failure, different things within standard 
theory can be blamed. Even the standard approach itself can be wrong if classical 
phenomena are the basic ones and quantum ones only emerge from it \cite{tHo99,Adl04,Wet11}.
In any case, a minimum lesson exists. We should extend
our investigations for all possible mathematical models of the mere coexistence
of the quantum and classical phenomena. I avoid talking about their interaction
because this term is reserved for something which is dynamical. Instead, I
talk about their mutual influence. Quantum and classical phenomena coexist where
coexistence includes their mutual dependence:
\begin{equation}
\mathrm{Classical}\Longleftrightarrow\mathrm{Quantum}
\end{equation}
Mathematically this means
that I allow for any coupling between the equation governing $\ro(t)$ of the
quantum sector and the equation governing $z(t)$ of the classical sector, respectively.     

The mathematical form of the influence of the classical on the quantum looks trivial. 
The Hamiltonian will be made $z$-dependent:
\begin{equation}
\frac{d\ro}{dt}=\frac{-\I}{\hbar}[\Ho(z),\ro]\;.
\end{equation}
This dependence may be time-local or time-non-local, it must respect 
causality anyway: $\Ho$ at time $t$ should not depend on $z(t')$ at $t'>t$.
 
The influence of quantum on classical, called back-reaction, is the problem.
In Sec.~\ref{Introduction} we mentioned that the measurement theory
can assign a certain $z$ to the current state $\ro$. It is also reassuring  
that the influence (\ref{coll}) of $z$ on $\ro$ is part of von Neumann theory. There is
a major defect: the von Neumann assignment of $z$ is instantaneous, it cannot
assign the continuum $z(t)$ to $\ro(t)$. The von Neumann
theory of measurement has been generalized for time-continuous measurements
so that in principle we possess a model of quantum-classical coexistence.
The realistic non-Markovian continuous measurement theory
is in its infancy, faces particular issues. 

\section{Five theories of coexistence}\label{Five}
Parallel to the measurement approach, many 
different---though necessarily interrelated---proposals have targeted the
quantum-classical coexistence and the implicit back-reaction. 
Some of them are very popular, some of them are less so.  
This time we ask the same question: do they pass FWT?

\subsection{Mean-field}
Robust quantum fields in quantum cosmology (also in nuclear, molecular, laser physics, e.t.c.) 
may look like classical fields. They can be replaced by
classical fields in mean-field theory \cite{Mol62,Ros63}. This is a very successful 
method to identify the classical continuum $z$ by the quantum mean value of the robust 
quantum field $\qo$ in question: 
\begin{equation}\label{z_meanf}
z=\tr(\qo\ro)\;.
\end{equation}
The mean-field $z(t)$ is smooth and causal.
To apply FWT, we make $\Ho(t)$ depend on $z(t)$:
\begin{equation}
\frac{d\ro}{dt}=\frac{-\I}{\hbar}[\Ho(z),\ro]\;.
\end{equation}
This means a non-linear evolution for $\ro$, it denies the statistical 
interpretation of $\ro(t)$. The mean-field $z$ is not tangible. Hence the
corresponding assignment $\ro\Rightarrow z$ may be a good approximation but 
it is flawed fundamentally (as it has already been known independently of FWT).

\subsection{De-Broglie---Bohm}
This is the oldest theory \cite{Boh52,DurTeu09} of assigning classical continuum to the quantum state. 
The assignment of the classical variable $z(t)$ concerns the coordinate operator $\qo$, the theory
is restricted for pure state density matrices $\ro=\ro^2$. The initial assignment $\ro(0)\rightarrow z(0)$ 
is random; its statistics is the same as if $\qo$ were measured in standard quantum measurement. 
This feature can be preserved by the following deterministic
evolution of the de-Broglie---Bohm trajectories:
\begin{equation}\label{z_Bohm}
m\frac{d^2 z}{dt^2}=-V'(z)-V'_{\ro}(z)\;.
\end{equation}
The classical continuum variable $z(t)$ senses 
the $\ro$-dependent quantum potential $V_{\ro}(z)$
in addition to the potential $V$ already present in $\Ho$.
The trajectory $z(t)$ is smooth and causal.

Does FWT pass? If we causally control $\Ho$ by $z$, e.g.: we add a deliberate potential $V(z)$, 
the theory remains consistent with quantum mechanics. If, however, we apply a more general
casual control and add a potential $V(z(t');t)$ at time $t$, controlled by $z(t')$ at an earlier
time $t'<t$, then the theory won't be consistent with standard quantum mechanics. 
The de-Broglie---Bohm trajectory $z$ is useful, e.g., in quantum chemistry calculations
\cite{PreBro01,WyaTra05} but it may prove not to be tangible, not to be fundamental assignment 
of classical continuum to the quantum system unless we confirm or ignore FWT. 

\subsection{Decoherence}\label{Decoherence}
This is the key mechanism \cite{Zeh71,Zur91,Giuetal03} of how quantum systems may, 
without measurement, express classical features. At certain
specific circumstances, the assignment $\ro\rightarrow z$ can be introduced even to
closed, unitarily evolving, quantum systems. 
A rigorous mathematical model \cite{Gri84,Omn92,GelHar93} can be based on the class (history-) 
operator \begin{equation}\label{Cz}
\Co_z=\Po^{(n)}_{z_n}\dots\Po^{(2)}_{z_2}\Po^{(1)}_{z_1},
\end{equation}
the time-ordered product of $n$ Hermitian projectors chosen from $n$ independent complete 
orthogonal sets. The candidate classical variable is the time-series 
$z=(z_1,z_2,\dots z_n)$ of the labels. The assignment is random, with
the probability $p_z=\tr(\Co_z^\dagger\Co_z\ro)$. The consistency of the
assignment requires the decoherence condition:
\begin{equation}
\tr(\Co_z^\dagger\Co_u\ro)=0~~~\mathrm{for}~z\neq u.
\end{equation}

Whether FWT passes or doesn't? We conjecture it does, at least under largely general circumstances.
The correct answer needs future investigations, see the proofs \cite{Dio04,Hal09} of a related 
simple test of dynamic robustness. 

If the classical variable $z$ turns out to be tangible still it needs a refinement  
since the construction (\ref{Cz}) is discrete,
the time-series of classical variables $z_1,z_2,\dots,z_n$ is discrete,
not a continuum yet.
To yield the tangible classical continuum $z(t)$ one has to take a
certain time-continuous limit of the theory. This limit is non-trivial,
it may just coincide \cite{DioGisHalPer95} with theories of time-continuous measurement. 

\subsection{Measurement} 
We learned in Sec.~\ref{Free} that a quantum measurement assigns a tangible
classical variable $z$ to the quantum state $\ro$. We can use sequential
measurements to assign the time-series $z_1,z_2,\dots,z_n$. The problem is
the same as with the decoherence theory. To assign a tangible 
classical continuum $z(t)$ one has to take the time-continuous limit.
 
The fundamental need of the continual assignment $\ro\Rightarrow z$ led
to the theory of time-continuous quantum measurement \cite{Dio88};
the theory was invented and fully developed independently in other important 
contexts \cite{Bel88,Car93,WisMil10}. Let's restrict ourselves for presenting the structure of
assignment itself:
\begin{equation}\label{z_whitenoise}
z=\tr(\qo\ro)+\mbox{white-noise}\;.
\end{equation}
This variable is the time-continuous outcome of the time-continuous measurement.
As such, it inherits the tangibility of single measurement outcomes: 
$z(t)$ can safely be used to control the Hamiltonian at any later time.
In other words, the assignment passes FWT. 
It is remarkable that the white-noise contribution 
turns the non-tangible mean-field (\ref{z_meanf}) into the tangible classical variable. 
The mathematics of time-continuous measurement theory re-appears in dynamical collapse theories 
\cite{Dio89,BasGir03}. Despite vast research on their ontological status, the tangible variable $z$ 
has traditionally been abandoned, with rare exceptions \cite{Dio89,DowHer05}. 

Unfortunately, we face a novel issue with the assignment (\ref{z_whitenoise}). 
The white-noise function is everywhere singular hence the
classical variable $z(t)$ becomes singular everywhere. The white-noise
is an artefact rooted in the simple memoryless (Markovian) model of
time-continuous measurement. We could get smooth colored noise and smooth 
tangible $z(t)$ in a non-Markovian theory \cite{StrDioGis99}:
\begin{equation}\label{z_colourednoise}
z=\tr(\qo\ro)+\mbox{colored-noise}\;.
\end{equation}
This brings causality issues in. Recent discussions predicts that
the tangible non-Markovian assignments $\ro\Rightarrow z$ are restricted by time-delay/retrodiction
\cite{Dio08a,WisGam08}. At the current time $t$ the assignment (\ref{z_colourednoise})
is tangible for earlier times $t'<t$ and $z(t')$ is not (yet) tangible at the current time $t$ 
if $t'$ gets closer to $t$ than the non-Markovian memory of the measurement.   

\subsection{Hybrid dynamics} 
When a classical continuum $z$ and a quantum system coexist we can assume 
dynamical coupling for them. We could just use an interaction Hamailtonian
$\Ho_I(z)$ which is a Hermitian operator as well as a function of the classical dynamical variable. 
For longtime it has been known that such purely canonical hybrid dynamics does 
not work. Nonetheless, there are various phenomenological constructions of hybrid 
dynamics and they describe the coexistence between classical continuum and quantum variables,
see Salcedo's summary \cite{Sal12} and references therein. 

Suppose a quantum system described by the density matrix $\ro$ coexists with a
classical system described by the density $\rho_C(z)$ over variables $z$ which
may be the canonical coordinates $\{z\}=\{(q,p)\}$. If the quantum and classical systems
have not yet influenced each other, i.e., they are independent, then their composite state 
can be described by the hybrid density $\ro(z)=\ro\rho_C(z)$. If we let them influence 
each other, the product structure is no longer valid. The evolution 
\begin{equation}
\ro\rho_C(z)\longrightarrow\ro(z;t)
\end{equation}
preserves positivity and normalization. The evolution must be linear
for the sake of statistical interpretation of $\ro(z;t)$ \cite{Dio11}.
This is the formal criterion of the statistical consistency discussed
by \cite{Sal12} in details. In proposals of hybrid dynamics of
ensemble coupling \cite{RegHal09} or coupling the mean-field to the classical 
sector, the hybrid state $\ro(z)$ would not follow a 
linear evolution, in accordance with the analysis in \cite{Sal12}. 
This does not necessarily mean the failure of FWT.

FWT is a different and more powerful test w.r.t. the statistical 
test in \cite{Sal12}. Some hybrid theories intend to replace or modify the standard 
statistical interpretation of the density matrix. In all these cases FWT requests 
that the given non-standard theory remain self-consistent under the $z$-dependent
causal control of the dynamics, no matter if $\ro(z;t)$ evolves
linearly or not, no matter if the hybrid dynamics is a sort of canonical
one or something radically different. However, the theories with
non-linear evolution of the hybrid density will easily loose their
self-consistence under the $z$-dependent causal control.

There are many versions of hybrid dynamics 
\cite{Dio95,DioHal98,DioGisStr00,ElzGamVal11,Mil12}
which are still to be checked against
FWT. Some are likely to pass, especially those which couple certain measurement
outcomes---tangible variables---to the classical sector, see \cite{DioHal98}
coupling the Markovian outcome (\ref{z_whitenoise}), \cite{DioGisStr00} coupling
the non-Markovian continuum of Husimi-variables. 

\section{Closing remarks}
The whole diversity of foundational theories of quantum-classical 
relationship, from measurement theory to hybrid dynamics, distills 
to the mathematical issue of coexistence of classical continuum and quantum variables.
This coexistence must respect certain basic constraints on one hand and
might be radically different from standard quantum theory on the other.
Only we have to formulate our basic constraints mathematically. This
concept was raised in \cite{Cam99,Bie04}. and further explained in \cite{Dio06}:
`In our longstanding struggles with the problem of Classical vs. Quantum, 
the main issue to overcome has always been the painful lack of a consistent
model that ``couples'' the coexisting classical and quantum entities. Aren't
quantum dynamics and measurement too restrictive? Are there any other consistent
mechanisms?'.

One of the non-trivial mathematical constraints is the Free Will Test.
It has nothing direct to do with the Free Will Theorem \cite{ConKoc06}.
A traditional definition of `free will' in quantum theory is our freedom to choose the
measurement setup, see e.g. in \cite{tHo07}. My definition is our free choice (control) of 
future dynamics (and measurement setups, if you like) in function of our knowledge obtained 
earlier. The deliberate $z$-dependent causal control of the dynamics will possess its 
mathematical representation within all plausible theory of classical-quantum coexistence. 
The condition that the theory remains self-consistent must be decidable in all theories. 
If not, it means that the theory should first be concretized.
That FWT may not be satisfied by a number of popular theories of
coexistence was also recognized in \cite{Cam99,Bie04}. (Tangible and non tangible 
classical variables were called just \emph{true} and \emph{not true} ones, respectively.)

I haven't considered the status of my proposal among current competing metaphyical concepts \cite{Cal09}.
It may become part of them or contribute to them in some form. We should not
overemphasize the metaphysical aspect at the expense of physics proper. The proposal 
is objective and mathematical, I could well replace the anthropic term
FWT by FCT: Free Control Test. It calls for direct checks (i.e.: calculations) in a 
number of theories, it shows the mathematical barriers against tangible classical
variables other than measurement outcomes in whatever sophisticated formalisms.    

\ack
This work was supported by the Hungarian Scientific Research Fund under Grant No. 75129,
the EU COST Action MP1006 `Fundamental Problems in Quantum Physics'. The author 
gratefully thanks the organizers of the International Conference on `Emergent Quantum Mechanics'.


\section*{References}
\begin{thebibliography}{99}
\bibitem{Neu32} von Neumann J 1932 {\it Mathematische Grundlagen der Quantenmechanik} (Berlin: Springer)
\bibitem{Dio11} Di\'osi 2011 L {\it Short Course in Quantum Information Theory} (Berlin: Springer)
\bibitem{Hus40} Husimi K 1940 {\it Proc. Phys. Math. Soc. Jpn.} {\bf 22} 264
\bibitem{tHo99} t' Hooft G 1999 {\it Class. Quant. Grav.} {\bf 16} 3263
\bibitem{Adl04} Adler S 2004 {\it Quantum Theory as an Emergent Phenomenon} (Cambridge: Cambridge University Press)
\bibitem{Wet11} Wetterich C 2011 {\it Preprint} arXiv:1201.6212 (this volume)

\bibitem{Mol62} Moller C 1962 in {\it Les Theories Relativistes de la Gravitation} ed  A Lichnerowicz and M.A. Tonnelat (Paris: CNRS)
\bibitem{Ros63} Rosenfeld L 1963 {\it Nucl. Phys.} {\bf 40} 353

\bibitem{Boh52} Bohm D 1952 {\it Phys. Rev.} {\bf 85} 166; {\it Phys. Rev.} {\bf 85} 180 
\bibitem{DurTeu09} D\"urr D and Teufel S 2009 {\it Bohmian Mechanics} (Berlin: Springer)
\bibitem{PreBro01} Prezhdo O V and Brooksby C 2001 {\it Phys. Rev. Lett.} {\bf 86} 3215 
\bibitem{WyaTra05} Wyatt R E and Trahan C J 2005 {\it Quantum dynamics with trajectories} (Berlin: Springer) 

\bibitem{Zeh71} Zeh H D 1971 in {\it Foundations of Quantum Mechanics} ed B. d'Espagnat (New York: Academic Press) p 263
\bibitem{Zur91} Zurek W 1991 {\it Phys. Today} {\bf 44} 36
\bibitem{Giuetal03} Giulini D, Joos E, Kiefer C, Kupsch J, Stamatescu I O and Zeh H D 2003 {\it Decoherence and the Appearance of a Classical World in Quantum Theory} (Berlin: Springer)
\bibitem{Gri84} Griffiths R B 1984 {\it J. Stat. Phys.} {\bf 36} 219
\bibitem{Omn92} Omn\`es R 1992 {\it Rev. Mod. Phys.} {\bf 64} 339
\bibitem{GelHar93} Gell-Mann M and Hartle J B 1993 {\it Phys. Rev.} {\bf D 47} 3345
\bibitem{Dio04} Di\'osi L 2004 {\it Phys. Rev. Lett.} {\bf 92} 170401
\bibitem{Hal09} Halliwell J J 2009 {\it Quant. Inf. Proc.} {\bf 8} 479
\bibitem{DioGisHalPer95} Di\'osi L, Gisin N, Halliwell J and Percival I C 1995 {\it Phys. Rev. Lett.} {\bf 74} 203

\bibitem{Dio88} Di\'osi L 1988 {\it Phys. Lett.} {\bf 129 A} 419 
\bibitem{Bel88} Belavkin V P 1989 {\it Phys. Lett.} {\bf 140 A} 355
\bibitem{Car93} Carmichael H 1993 {\it An Open Systems Approach to Quantum Optics} (Berlin: Springer)
\bibitem{WisMil10} H.M. Wiseman H M and Milburn G J 2010 {\it Quantum Measurement and Control} (Cambridge: Cambridge University Press) 
\bibitem{Dio89} Di\'osi L 1989 {\it Phys. Rev.} {\bf A 40} 1165
\bibitem{BasGir03} Bassi A and Ghirardi G C 2003 {\it Physics Reports} {\bf 379} 257
\bibitem{DowHer05} Dowker F and Herbauts I 2005 {\it Found. Phys. Lett.} {\bf 18} 499 
\bibitem{StrDioGis99} Strunz W T, Di\'osi L and Gisin N 1999 {\it Phys. Rev. Lett.} {\bf 82} 1801
\bibitem{Dio08a} Di\'osi L 2008 {\it Phys. Rev. Lett.} {\bf 100} 080401; {\bf 101} 149902(E)
\bibitem{WisGam08} Wiseman H M and Gambetta J M 2008 {\it Phys. Rev. Lett.} {\bf 101} 140401

\bibitem{Sal12} Salcedo L L 2012 Statistical consistency of quantum-classical hybrids  {\it Preprint} arXiv:1201.4237
\bibitem{RegHal09} Reginatto M and Hall J W  2009 {\it J. Phys.: Conf. Ser.} {\bf 174} 012038
\bibitem{Dio95} Di\'osi L 1995 Quantum dynamics with two Planck constants and the semiclassical limit {\it Preprint} arXiv:quant-ph/9503023
\bibitem{DioHal98} Di\'osi L and Halliwell J J 1998 {\it Phys. Rev. Lett.} {\bf 81} 2846 
\bibitem{DioGisStr00} Di\'osi L, Gisin N and Strunz W T 2000 \textit{Phys. Rev.} {\bf A 61} 22108
\bibitem{ElzGamVal11} Elze H T, Gambarotta G and Vallone F 2011 {\it J. Phys. Conf. Ser.} {\bf 306} 012010 
\bibitem{Mil12} Milburn G J 2012 Decoherence and the conditions for the classical control of quantum systems {\it Preprint} arXiv:1201.5111

\bibitem{Cam99} Di\'osi L 1999 {\it Talk at} 'Complexity, Computation and the Physics of Information' (Cambridge, 5-23 July 1999) www.rmki.kfki.hu/\~{}diosi/slides/cambridge.pdf
\bibitem{Bie04} Di\'osi L 2004 {\it Talk at} 'Quantum Theory Without Observers II' (Bielefeld, 2-6 February 2004) www.rmki.kfki.hu/\~{}diosi/slides/bielefeld.pdf
\bibitem{Dio06} Di\'osi L 2006 {\it AIP Conf. Proc.} {\bf 844} 133

\bibitem{ConKoc06} Conway J and Kochen S 2006 {\it Found. Phys.} {\bf 36} 1441 
\bibitem{tHo07} t' Hooft G 2007 The Free-Will Postulate in Quantum Mechanics {\it Preprint} arXiv:quant-ph/0701097
\bibitem{Cal09} Callender C 2009 in {\it Compendium of Quantum Physics} ed Greenberger D M, Hentschel K and Weinert F (Berlin: Springer) p 384

\end{thebibliography}
\end{document}